\documentclass[usenatbib,useAMS]{mn2e}

\usepackage{natbib}
\usepackage{graphicx,epsfig}
\usepackage{latexsym,amssymb}
\usepackage[fleqn]{amsmath}
\usepackage{color}

\newcommand{\hkpc}{$h^{-1}$\,kpc}
\newcommand{\hmpc}{$h^{-1}$\,Mpc}

\newcommand{\beq}{\begin{equation}}
\newcommand{\eeq}{\end{equation}}
\newcommand{\beqa}{\begin{eqnarray}}
\newcommand{\eeqa}{\end{eqnarray}}
\newcommand{\rmd}{\ensuremath{\mathrm{d}}}

\newcommand{\ds}{\ensuremath{\Delta\Sigma}}

\title[Testing adiabatic contraction]{Testing adiabatic contraction
  with SDSS elliptical galaxies}

\author[Schulz et al.]{%
  A.E. Schulz$^1$\thanks{\texttt{schulz@ias.edu}},
  Rachel Mandelbaum$^2$, 
  Nikhil Padmanabhan$^3$
  \\
  $^1$Institute for Advanced Study, Einstein Drive, Princeton NJ
  08540, USA 
  \\
  $^2$Department of Astrophysical Sciences, Princeton University, Peyton Hall, Princeton, NJ 08544, USA
  \\
  $^3$Department of Physics, Yale University, New Haven, CT 06511
}

\begin{document}

\date{\today}

\maketitle 

\begin{abstract}
  We study the profiles of 75~086 elliptical galaxies from the Sloan Digital Sky
  Survey (SDSS) at both large ($70-700$ $h_{70}^{-1}$kpc) 
  and small ($\sim 4$ $h_{70}^{-1}$kpc) scales.  Weak lensing observations
  in the outskirts of the halo are combined with 
  measurements of the stellar velocity dispersion 
  in the interior regions of the galaxy for stacked galaxy samples.
  The weak lensing measurements
  are well characterized by a Navarro, Frenk and White (NFW) profile. 
  The dynamical mass measurements exceed the extrapolated NFW profile
even after the estimated stellar masses are subtracted, providing
evidence for the
modification of the dark matter profile by the baryons. This excess
mass is quantitatively consistent with the predictions of the adiabatic contraction (AC)
hypothesis.
Our finding suggests that
  the effects of AC during galaxy formation are
  stable to subsequent bombardment from major and minor mergers.
  We explore several theoretical and observational
  systematics and conclude that they cannot account for the inferred
  mass excess.  
  The most significant source of systematic error is in the IMF, 
  which would have to increase the stellar mass 
  estimates by a factor of two relative to the Kroupa IMF to fully explain 
  the mass excess without AC, causing 
 tension with results from SAURON \citep{2006MNRAS.366.1126C}.  
  We demonstrate a connection between the level of 
  contraction of the dark matter halo profile and 
  scatter in the size luminosity relation, which is a projection of the fundamental plane.
  Whether or not AC is the mechanism supplying the
  excess mass, models of galaxy formation and evolution must reconcile the observed
  halo masses from weak lensing with the comparatively large dynamical
  masses at the half light radii of the galaxies. 
  
\end{abstract}


\section{Introduction}\label{sec:intro}
  In hierarchical models of structure formation, matter organizes itself into
gravitationally bound halos that are formed through mergers of smaller halos, and 
by accretion along filamentary large scale structure.  The properties of these 
halos have been studied extensively in N-body Cold Dark Matter (CDM) simulations.  Dark 
matter halos have been shown to exhibit a universal density profile, of which the most common 
parametrization is referred to as
the NFW profile \citep{1997ApJ...490..493N}.  
More recent studies have determined that dark matter profiles 
deviate on average slightly from this two parameter description, and are better described by 
the Einasto profile, which introduces one extra parameter that mitigates the tendency 
for the measured concentration to spuriously depend on the innermost radius in the fit to the density profile
(\citealt{2004MNRAS.349.1039N,2005ApJ...624L..85M,2008MNRAS.387..536G}). 
The regularity in dark matter halos is one of the most robust predictions 
of hierarchical structure formation. 

Galaxies are believed to form inside dark matter halos
via the cooling and condensation of baryons to their centers.  Hydrodynamic 
simulations of hot gas in halos indicate that as this process occurs, baryonic physics
affects the shape of the dark matter profile.  As the baryons cool to the 
center, they begin to dominate the mass in the halo interior.  The dark matter feels 
the change in the gravitational potential and is drawn into the center, steepening the 
central profile.   The response of the dark matter to the cooling baryons has 
been modeled as the  adiabatic contraction (AC) of a series of spherical shells under the 
assumptions of spherical symmetry, circular particle orbits, and conservation of angular 
momentum \citep{1986ApJ...301...27B}.  This treatment has been modified to account for the orbital 
eccentricity of particles in \cite{2004ApJ...616...16G}, and has been shown to reproduce the profiles
of halos in hydrodynamical simulations at the 10--20 per cent level.

Realistically, the AC process is synchronous with other more 
violent processes of merging and accretion.  Older galaxies, in which
the gas has been 
converted to stars long ago, continue to suffer dissipationless 
merging that has been argued to remove the influence of AC 
\citep{2004ApJ...614...17G,2009ApJ...697L..38J}.  If 
a galaxy has had some portion of its baryonic content 
shock heated and removed to the warm-hot intergalactic medium where it is no longer 
seen, then potentially the contraction of the dark matter halo will be smaller, while the halo 
mass will remain the same.  The hydrodynamic simulations of \cite{2008ApJ...672...19R} 
suggest that the AC of the halo persists at a level that affects the matter power spectrum at 
halo-sized scales.  On the other hand, \cite{2007ApJ...658..710N}
suggest that the galaxy formation history, and specifically whether
recent growth is due to star formation versus due to mergers 
and/or accretion, influences whether adiabatic contraction occurs or 
whether the dark matter halo in fact becomes less concentrated.  Due to the many theoretical
uncertainties, an observational handle on the mass distribution in
both the inner and outer regions of the halo is crucial. 

The most reliable way to probe the 
dark matter is through its gravitational effects.  
%
Weak gravitational lensing (for a review,
see \citealt{2001PhR...340..291B}) of background galaxies by a
foreground dark matter halo, or galaxy-galaxy lensing, is one robust way
to constrain the halo density profile. This method has been used to
constrain halo density profiles outside of the regions where the
influence of baryons is significant, for stacked galaxy samples (due to
the lower signal) and for stacked and individual clusters
(e.g. \citealt{2003ApJ...588L..73D,2006MNRAS.372..758M,2008JCAP...08..006M,2009arXiv0903.1103O,2009ApJ...693.1570R}).  However, the inner regions of the halo where
AC is important are typically inaccessible to weak lensing except perhaps
for very deep space-based observations (e.g.,
\citealt{2007ApJ...667..176G} measure the weak lensing signal down to 10
\hkpc).  In this regime, the number of available background galaxies is small, confusion
of light from the lens galaxy can make the measurement of background
galaxy shapes unreliable, and the weak lensing approximation may fail
entirely on small scales.  Progress can be made 
by combining weak lensing with another probe
such as strong lensing or kinematics 
\citep[e.g.,][]{2003ApJ...598..804K,2007ApJ...667..176G,2007ApJ...668..643L,
2008ApJ...681..187B,2009MNRAS.393.1235C,2009ApJ...699.1038O,
2009ApJ...694.1643U}.

In this paper, we combine weak lensing observations in
the halo, outside of the central regions ($>70$ $h_{70}^{-1}$kpc),
with dynamical measurements
that probe the central regions ($<10$ $h_{70}^{-1}$kpc) of a stacked sample of elliptical galaxies 
from the Sloan Digital Sky Survey (SDSS).   
This approach allows us to constrain the 
density profile over a large dynamic range of scales. An alternate
approach is to combine strong and weak
  lensing measurements, as in \cite{2007ApJ...667..176G}; however, our
  approach allows the use of a statistical sample of elliptical galaxies,
  rather than being constrained only to those that are strong lenses.
  Finally, integral-field spectroscopy \citep[e.g., ][]{2002MNRAS.329..513D,2006MNRAS.366.1126C} can be used
  to probe the galaxy density velocity distribution and therefore
  infer the matter profile in detail within the effective radius,
  which in comparison with stellar population synthesis models applied to the
  light distribution, can be used to infer the amount of dark matter
  within the effective radius.  However, the connection to the profile
  on much larger scales is unclear, so it is impossible on the basis
  of IFU observations alone to distinguish between an adiabatically
  contracted halo versus a halo that is very massive overall.

In section \ref{sec:obs}, we describe the sample of galaxies used in the analysis. 
In section \ref{sec:results}, we begin with the dynamical mass measurement and then combine 
it with weak lensing observations.  We further divide the sample along the mean size-luminosity 
relation and show that the scatter from this relation correlates with properties of the underlying 
dark matter profile.  We conclude in section \ref{sec:conclusions}.  Throughout, we use comoving 
distances unless specifically noted, with lowercase $r$ representing the three dimensional
separation and uppercase $R$ representing projected
quantities.  We have assumed a flat LCDM cosmology with $\Omega_m=0.3$ and
$\Omega_\Lambda=0.7$. Since we must compare stellar masses, which
scale like $h^{-2}$ (for Hubble parameter $H_0=100 h$km/s/Mpc), with
dynamical and halo masses, which scale like $h^{-1}$, our results are
by necessity dependent on the choice of $h$.  We use $h=0.7$ for all
quantities, and explicitly indicate their $h$-dependence using $h_{70}$;
e.g., halo masses are indicated with $h_{70}^{-1}M_{\odot}$, stellar
masses with $h_{70}^{-2}M_{\odot}$, distances with $h_{70}^{-1}$kpc, ratios of stellar to dynamical
mass with $h_{70}^{-1}$, and similarly for other quantities.  When relating results from
other works that use $h=1$, we simply use \hkpc\ or
$h^{-1}M_{\odot}$.  

\section{Observational Sample}\label{sec:obs}

The lens and source samples that we use are from the SDSS
\citep{2000AJ....120.1579Y}, which has imaged roughly $\pi$ steradians
of the sky, and followed up approximately one million of the detected
objects spectroscopically \citep{2001AJ....122.2267E,
  2002AJ....123.2945R,2002AJ....124.1810S}. The imaging was carried
out by drift-scanning the sky in photometric conditions
\citep{2001AJ....122.2129H, 2004AN....325..583I}, in five bands
($ugriz$) \citep{1996AJ....111.1748F, 2002AJ....123.2121S} using a
specially-designed wide-field camera
\citep{1998AJ....116.3040G}.   All of
the data were processed by completely automated pipelines that detect
and measure photometric properties of objects, and astrometrically
calibrate the data \citep{2001adass..10..269L,
  2003AJ....125.1559P,2006AN....327..821T,2008ApJ...674.1217P}. The SDSS was completed
with its seventh data release, DR7  
\citep{2002AJ....123..485S,
  2003AJ....126.2081A, 
  2004AJ....128..502A, 
  2005AJ....129.1755A,
  2004AJ....128.2577F,
  2006ApJS..162...38A,
  2007ApJS..172..634A,
  2008ApJS..175..297A,
  2009ApJS..182..543A}, from which we derive our sample, though it
does not cover the entire area.


\subsection{Lens sample}
The
lens galaxies are taken from the spectroscopic sample, for which we
used the NYU Value Added Galaxy Catalog 
(VAGC, \citealt{2005AJ....129.2562B}).  For the portion of the analysis that requires 
stellar masses, we have used the stellar masses of
\cite{2003MNRAS.341...33K} updated for
DR7\footnote{www.mpa-garching.mpg.de/SDSS/DR7/Data/stellarmass.html}.

The lens population is selected from the SDSS spectroscopic sample
using the following criteria.  
These criteria are designed to select elliptical or S0 type galaxies with little or no
rotational support that are the central galaxy in their dark matter halos, so that SDSS velocity dispersion measurements can be reliably 
translated into dynamical masses, and the weak lensing signal more easily interpreted.  All quantities are from the standard SDSS
reductions (rerun 137), unless specifically mentioned.
\begin{itemize}
\item $0.02<z<0.35$: We select lenses from the Main spectroscopic sample with these spectroscopic redshifts. 
\item $g-r > 0.8$ : This cut in rest-frame colour ensures that we are looking only at red galaxies 
with little to no active star formation.   We have used model magnitudes $k$-corrected to redshift
$0.1$ from the VAGC \citep{2007AJ....133..734B}.  The model magnitudes are extinction corrected using the reddening maps of \cite{1998ApJ...500..525S} and extinction-to-reddening ratios from
\citet{2002AJ....123..485S}.
\item $\sigma_v > 70$ km/sec : This cut ensures that the stellar velocity dispersion in the SDSS fiber 
can be reliably determined, given the resolution of the SDSS spectrograph.
\item $R_{90}/R_{50} > 2.6$: This cut on the ratio of the Petrosian $r$-band
  90 and 50 per cent light radii helps to ensure that we are selecting
elliptical lens galaxies.  The fact that the intensity profiles of
ellipticals are more concentrated than those of spiral galaxies 
\citep{2001AJ....122.1238S} motivates this criterion. 
\item $b/a>0.7$: We select objects with large values of the minor to
  major semi-axis ratio, obtained from model fits to a de Vaucouleurs
  profile. This cut will eliminate large edge-on spiral galaxies
  (which may pass our colour cuts due to internal reddening), as well as elliptical galaxies with substantial rotation velocities that tend to flatten them.
\item Central: To ensure the lens galaxy is the central galaxy in its dark matter halo rather than 
a satellite galaxy, we assume that the brightest galaxy in a halo will
be at the center.  Thus we discard galaxies that have a brighter
neighbor within a cylinder centered on the galaxy.  Following
\cite{2009ApJ...698..143R}, we select a cylinder of diameter
1.6 \hmpc\ (using $h=1$ here for consistency with that work) and $\pm$ 864 km/s along the line of sight.
Due to fiber collisions in the SDSS, some of the galaxies that pass the cuts imposed on the Main spectroscopic 
sample lack redshifts.  These galaxies are not included in our lens catalog, but for 
the purposes of finding brighter neighbors, the redshift of the colliding galaxy is assigned 
to the object with no redshift to decide whether it falls into the cylinder.  
\end{itemize}
In addition to the selection made on the basis of lens galaxy
properties, we also removed galaxies that lacked source
galaxies in the background, that lacked stellar mass estimates, and that fell outside the mask of the
random catalogs available in the VAGC.  These cuts eliminated
approximately 15 per cent of the galaxies.

\begin{figure}
\begin{center}
\resizebox{3.6 in}{!}{\includegraphics{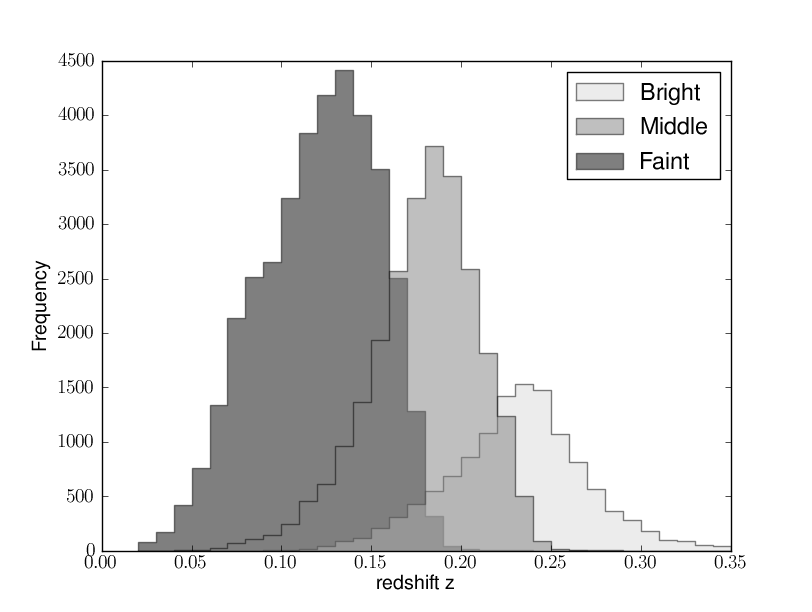}}
\end{center}
\caption{The redshift distribution of the the three luminosity bins
  used in this analysis.}
\label{fig:zhist}
\end{figure}

\begin{figure}
\begin{center}
\resizebox{3.6 in}{!}{\includegraphics{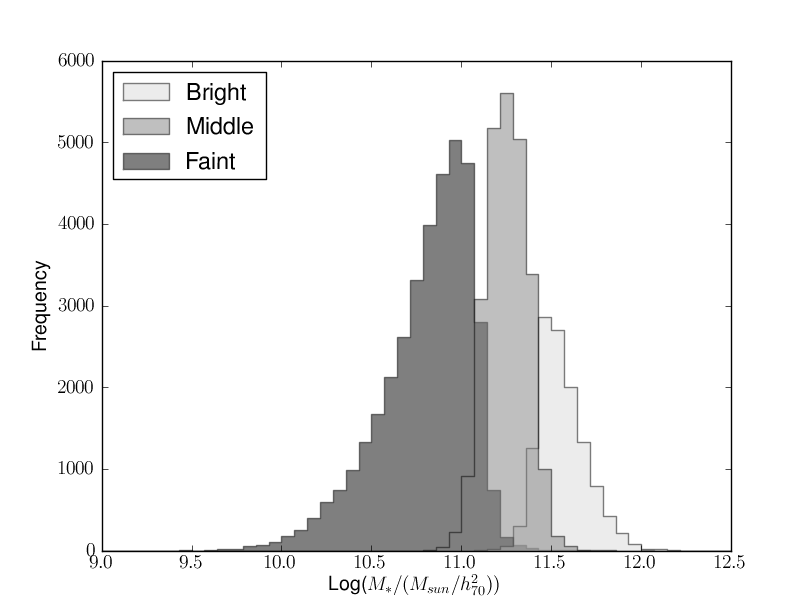}}
\end{center}
\caption{The distribution of stellar masses.}
\label{fig:stmh}
\end{figure}

\begin{figure}
\begin{center}
\resizebox{3.5 in}{!}{\includegraphics{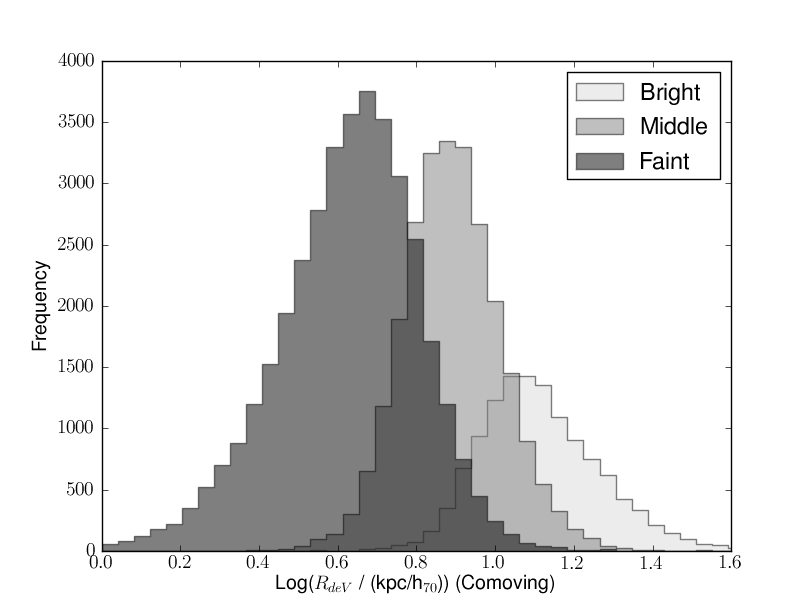}}
\end{center}
\caption{A histogram of the (comoving) projected half light radii for each of the three
luminosity samples. }
\label{fig:rhist}
\end{figure}

For the bulk of our analysis, we have divided the 75~086 lenses into three luminosity bins.  
The faintest bin contains half the lenses, the middle bin contains 1/3, and the brightest bin 
contains 1/6 of the sample to roughly
balance the signal-to-noise ratio ($S/N$) in the weak lensing
measurements.  Figure \ref{fig:zhist} shows the redshift distribution
for each of these bins, and Fig. \ref{fig:stmh} shows the
distribution of stellar masses. For each galaxy, we use the median of the PDF of the 
stellar mass as our stellar mass estimate. A histogram of the comoving half light 
radii is plotted in Fig. \ref{fig:rhist}

\subsection{Source sample}

The source galaxies (with shape estimates) are taken from the 
SDSS photometric catalogs, with additional processing and selection cuts
described in \cite{2005MNRAS.361.1287M}.  
This source
sample contains over 30 million galaxies from the SDSS imaging data with
$r$-band model magnitude brighter than 21.8, with shape measurements
obtained using the REGLENS pipeline, including PSF correction done via
re-Gaussianization \citep{2003MNRAS.343..459H}, and with cuts on
apparent size relative to the PSF designed
to avoid various shear calibration biases.  The overall calibration
uncertainty due to all systematics was originally estimated to be
eight per cent \citep{2005MNRAS.361.1287M}, though the redshift
calibration component of this systematic error budget has recently
been decreased due to the availability of more spectroscopic data
\citep{2008MNRAS.386..781M}.

For the lensing analysis, a subsample of the total source catalog is
used in probing the smaller radii, $R<100$\hkpc\ (described in section
\ref{sec:lensing}).  This procedure is a result of the analysis in
\cite{2006MNRAS.372..758M}, which found that for bright elliptical
lenses, use of the full source sample led to inclusion of satellites
of those lenses that were intrinsically aligned towards their hosts,
thus contaminating the lensing measurement.  The LRG sources used at
small separations were
identified using colour cuts described in \cite{2005MNRAS.361.1287M} to
have very little contamination from galaxies below $z=0.4$, thus
minimizing contamination from intrinsic alignments with our lenses, which
are well below this redshift.

\section{Results}\label{sec:results}
Table \ref{tab:results} summarizes all the properties of the lens sample used in this work.  We 
now describe how we arrived at these numbers. 
\begin{table*}
\centering
\begin{tabular}{|c|c|c|c|c|c|c|c|c|c|c|c|}
Luminosity&$\!\!$Sample$\!\!$&$z$&$\!\!\!\!M^{\rm Tot}_* (\rm e10)\!\!\!\!$&$\!\!M (\rm e13)\!\!$&c&$\!\!R_{\rm deV}\!\!$&$\!\!M^{\rm Tot}_{\rm dy} (\rm e10)\!\!$&$M^{\rm DM}_{\rm dy} (\rm e10)$&$M_{\rm nfw} (\rm e10)$&$f_*$   \vspace{0.2cm} \\ 
 & & &$h^{-2}_{70} M_\odot$ &$h^{-1}_{70} M_\odot$ & &$h^{-1}_{70}$ kpc& $h^{-1}_{70} M_\odot$ &$h^{-1}_{70} M_\odot$  &$h^{-1}_{70} M_\odot $ & $h^{-1}_{70}$ & \\
\hline
$M_r < -21.27$                 & Full &$0.22$ & $36 \pm 6$ & $5.7_{-0.8}^{+0.9}$ & $8_{-1.4}^{+1.6}$  & $13.4 \pm 0.5$ & $49  \pm    5$ & $34  \pm 6$ & $19_{-3}^{+4}$ &$0.31$ \vspace{0.2cm}\\ \vspace{0.2cm}
$-21.27< M_r < -20.54$ & Full&$0.17$ & $18 \pm 4$ & $2.4_{-0.3}^{+0.4}$ & $7_{-1.0}^{+1.4}$  & $8.0   \pm 0.4$ & $21  \pm    2$ & $14   \pm   2$ & $4.0_{-0.6}^{+0.9}$ &$0.35$ \\
$M_r > -20.54$                 & Full &$0.11$ &$7   \pm 2$ & $0.7_{-0.1}^{+0.2}$ & $9_{-2.2}^{+4.0}$  & $4.3   \pm 0.3$ & $7.3 \pm 0.7$ & $ 4.5  \pm 1.0$ & $1.1 _{-0.3}^{+0.6}$ &$0.38$  \\ 
\hline
$M_r < -21.27$                 &S/B & $0.22$ & $33 \pm 5$ & $3.0_{-0.8}^{+0.9}$ & $8_{-2}^{+3}  $     & $10.0 \pm 0.4$ & $39 \pm 3 $ & $25 \pm 4$ & $8_{-2}^{+4}$ & $0.36$   \vspace{0.2cm} \\ \vspace{0.2cm}
$M_r < -21.27$                 &L/D & $0.22$ & $38 \pm 6$ & $6.2_{-1.1}^{+1.3}$ & $9_{-2}^{+2}  $     & $16.1 \pm 0.6$ & $57 \pm 6 $ & $41 \pm 7$ & $30_{-6}^{+7}$ & $0.28$  \\ \vspace{0.2cm}
$-21.27 < M_r < -20.54$ &S/B & $0.17$ & $18 \pm 5$ & $1.9_{-0.4}^{+0.5}$ & $6_{-2}^{+2}  $     & $  6.5 \pm 0.3$ & $19 \pm 1$ & $11 \pm 2 $ & $2.2_{-0.6}^{+0.9}$ & $0.39$  \\ \vspace{0.2cm}
$-21.27 < M_r < -20.54$ &L/D & $0.17$ & $18 \pm 3$ & $2.8_{-0.6}^{+0.8}$ & $6_{-2}^{+2}  $     & $  9.8 \pm 0.4$ & $24 \pm 2$ & $17 \pm 2 $ & $5.6_{-1.4}^{+1.8}$ & $0.32$  \\ \vspace{0.2cm}
$M_r > -20.54$                 &S/B & $0.12$ & $  7 \pm 2$ & $0.6_{-0.2}^{+0.3}$ & $8_{-4}^{+7}  $     & $  3.5 \pm 0.2$ & $6.7 \pm 0.6$ & $3.9 \pm 0.9$ & $0.6_{-0.2}^{+0.6}$ & $0.42$  \\ \vspace{0.2cm}
$M_r > -20.54$                 &L/D & $0.11$ & $  7 \pm 2$ & $0.6_{-0.2}^{+0.3}$ & $13_{-5}^{+12}$  & $  5.2 \pm 0.3$ & $7.9 \pm 0.8$ & $5.2 \pm 1.0$ & $2.2_{-1.0}^{+2.8}$ & $0.35$  \\

\end{tabular}
\caption{Properties of the three lens samples studied in this work.  The upper part of the table shows the result for
the full sample, and the lower part shows the division of the sample along the mean size-luminosity relation (Smaller/Brighter
or Larger/Dimmer than the mean relation).  The notation e$10$ is shorthand for $(\times 10^{10})$. 
Luminosities are in absolute $r$-band magnitude, 
galactic extinction- and $k$-corrected as described in section \ref{sec:obs}.  The quantities $M$ and $c$ are fit
to the weak lensing observations.  The quantities $z$, $M_*$, $R_{\rm deV}$ (comoving coordinates)
and DM $M_\mathrm{dy}$ are lensing weighted averages over the lenses in the sample, and the Total 
$M_\mathrm{dy}$ is obtained by adding $0.42M_*$ to the DM $M_\mathrm{dy}$.  $M_{\rm nfw}$ is the mass expected 
from an NFW profile with mass $M$ and concentration $c$.  The dynamical masses and $M_{\rm nfw}$ are measured 
inside $R_{\rm deV}$.  The value $h=0.7$ has been inserted into the distance modulus.  
The fraction of mass contributed by the stars at $R_{\rm deV}$ is given by $f_*$, and depends on the value of h.   
Errors on $M$, $c$, and $M_{\rm nfw}$ are the 68 per cent CL obtained from bootstrap resampling of the NFW fits.  Errors on $M^{\rm Tot}_*$
and $R_{\rm deV}$  are measurement errors.   Errors on the dynamical masses come from a 10 per cent estimate of 
systematic error in the mapping between the velocity dispersion and the mass interior to $R_{\rm deV}$.}
\label{tab:results}
\end{table*}

\subsection{Dynamical mass measurements}\label{sec:dym}
We measure the dynamical mass at the 2D de Vaucouleurs scale radius
for comparison with the weak lensing 
dark matter profile measurement, described later in section \ref{sec:lensing}.  The dynamical mass is
inferred by assuming a relation between the velocity dispersion of stars inside the portion of the galaxy 
covered by the SDSS fiber (3 arcsec in diameter), and the mass enclosed within the 2D de Vaucouleurs scale radius. 
The mass is related to the velocity of particles on circular orbits as
\begin{align}\label{eqn:dym}
v_c^2(r_{\rm phys})=\frac{GM(r_{\rm phys})}{r_{\rm phys}},
\end{align}
but relating the circular velocity to the radial velocity dispersion $\sigma$ measured in SDSS requires a model 
for the stellar and dark matter mass distributions. The method for relating the two via the Jeans 
equation can be found in 
\cite{2004NewA....9..329P}.    We have assumed the stars follow a Hernquist profile 
\citep{1990ApJ...356..359H} and the dark matter profile is
an adiabatically contracted NFW profile, using the model of \cite{2004ApJ...616...16G}.
Since the sample is comprised of early type elliptical galaxies, we also assume that star formation 
has used all the 
available gas 
(\citealt{2005RSPTA.363.2693R}, \citealt{2006MNRAS.371..157M}, \citealt{1998ApJ...503..518F}) and 
we neglect any contribution to the mass from gas in the central region. 
We assume that there is no radial anisotropy
$\beta$ in the Jeans equation defined as 
\begin{align}
\beta=1-\frac{\bar{v}_t^2}{\bar{v}_r^2} =0
\end{align}
where $\bar{v}_r^2$ is the radial velocity dispersion 
and $\bar{v}_t^2$ is the tangential velocity dispersion, which is not observable.  Our results are insensitive to this assumption, as discussed 
section \ref{sec:se}.

Fig. 3 compares the
theoretical expectations for the velocity dispersion and
circular velocity profiles for our middle luminosity bin. The velocity
dispersion profile changes by $< 4$ per cent from $r/R_{\rm deV} = 0.5$ to
$2.5$, allowing us to extrapolate the velocity dispersion measurements
from the fiber radius to the half-light radius. Furthermore,
we find that the circular velocity is proportional to the velocity
dispersion at the $5$ per cent level, allowing us to robustly convert
velocity dispersions into dynamical masses. At the projected half
light radius, we find $v_c = 1.7 \sigma$.
Thus we can invert Eq.~\ref{eqn:dym} to solve for the dynamical mass.
\begin{align}\label{eqn:dym2}
M_\mathrm{dy}=\frac{(1.7 \sigma)^2 \, R_{\rm deV}}{G (1+z)}
\end{align}
Here we explicitly divide by $1+z$ since our notation has $R_{\rm deV}$ in comoving coordinates. 
Fig. 2 of \cite{2004NewA....9..329P}  shows that these curves are extremely insensitive to the value of the 
radial anisotropy parameter $\beta$, even if $\beta$ is not held constant with radius.
The constant of proportionality is also  insensitive (at $\sim 5$ per cent) to 
the stellar and DM mass profiles, 
and the details of the AC prescription, including no AC  \citep{2004NewA....9..329P} . 
Variations in these model quantities cause 
$M_\mathrm{dy}$ to shift at the level of $\sim 10$ per cent.   We adopt
this as the systematic uncertainty on the measurement of the total dynamical mass inside the SDSS fiber.
\begin{figure}
\begin{center}
\resizebox{3.5 in}{!}{\includegraphics{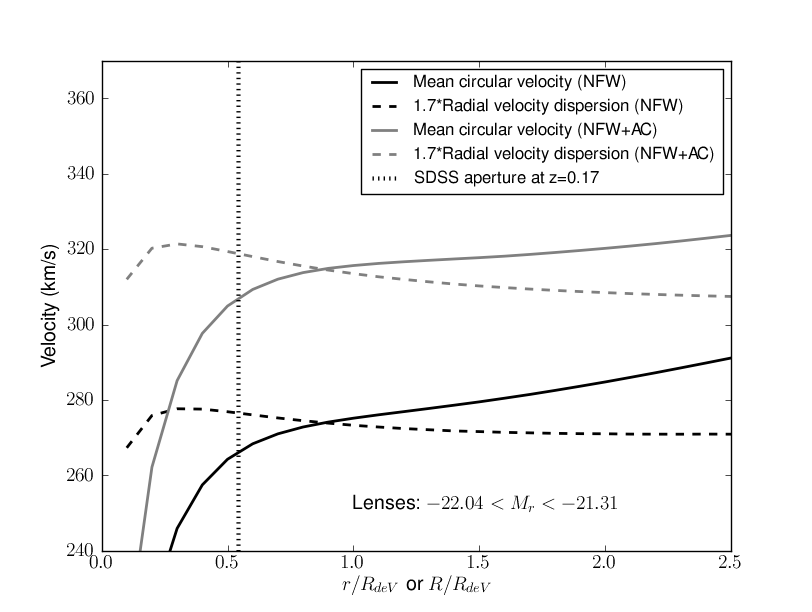}}
\end{center}
\caption{The circular velocity profile and radial velocity 
dispersion multiplied by 1.7, which causes the two to agree near the scale radius.  The 
parameters used to generate these curves are taken from the intermediate lens 
sample.  Note that these 
are relatively flat functions over a wide range of radii.  The vertical line marks the 
angular size of the SDSS fiber at the mean redshift of the middle luminosity sample.}
\label{fig:vel}
\end{figure}

Notice that the velocity dispersion $\sigma$ measured in SDSS is not measured at 
the projected half light radius $R_{\rm deV}$, but rather at the radius subtended by the SDSS fiber.  We 
appeal to the flatness of $\sigma(r)$ demonstrated in Fig. \ref{fig:vel} and assume that this measurement will
equal the velocity dispersion at the half light radius.  The angular diameter subtended by the SDSS fiber 
at the redshift of the middle luminosity bin is marked with a vertical line on the plot.  

\begin{figure}
\begin{center}
\resizebox{3.5 in}{!}{\includegraphics{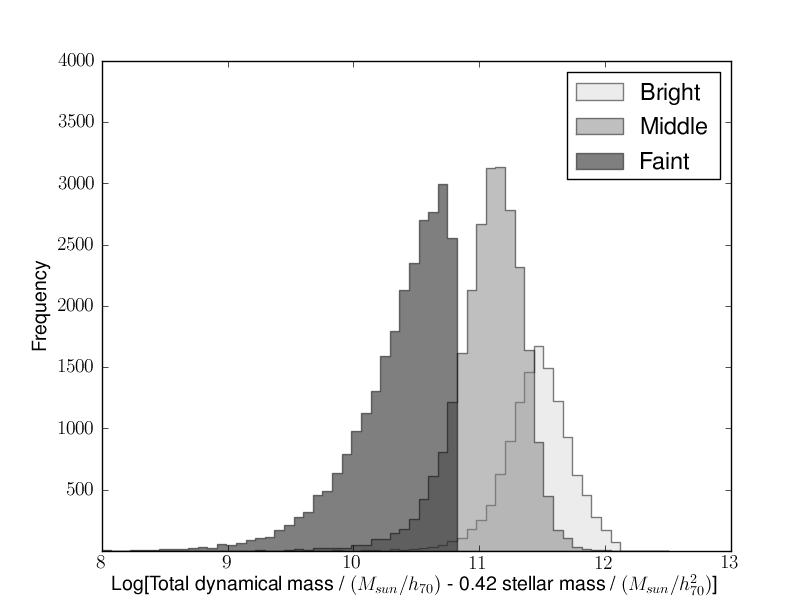}}
\end{center}
\caption{A histogram of the dark matter contribution to the dynamical mass for each 
of the three luminosity samples. }
\label{fig:dmdymhist}
\end{figure}

We emphasize one detail: although the radius
in Eq.~\ref{eqn:dym2} is the 2D projected half light radius, which is measured in SDSS by fitting to
a de Vaucouleurs profile, the dynamical mass in Eq.~\ref{eqn:dym2} is an estimate of the 
3D mass interior to that distance, not the mass in a 2D cylindrical projection through the halo out to that radius. 
Eq.~\ref{eqn:dym2} implies that at these radii the 3D dynamical mass is simply 
proportional to the radius (i.e., the profile is effectively isothermal), and we are choosing to make the measurement at the half-light radius
in order to be able to estimate the stellar contribution to the dynamical mass.   The 3D stellar mass 
contained within the 2D projected half light radius $R_{\rm deV}$ can be derived from the Hernquist 
profile to be roughly 42 per cent of the total stellar mass.  We subtract this 
from the total dynamical mass to derive the dark matter contribution 
to the dynamical mass at the half light radius.  A histogram of this quantity is plotted in 
Fig \ref{fig:dmdymhist}.  

\subsection{Combining with weak lensing measurements}\label{sec:lensing}

Weak gravitational lensing probes the differential surface density of the 
dark matter associated with the lens galaxies. On scales probed here, 
this surface density can be attributed to
the dark matter halo; on larger scales, due to large-scale structure.
Thus, on the scales less than $\sim 1$ $h_{70}^{-1}$Mpc, we can define the
projected surface density via
\begin{equation}\label{eqn:proj}
\Sigma(R)=\int \rho(R,z) \, \mathrm{d}z.
\end{equation}
Here $\rho(r=\sqrt{R^2+z^2})$ is the 3D density profile of the lens
galaxies. 

Weak lensing is sensitive to the differential surface density, defined as
\begin{equation}\label{eqn:ds}
\ds(R)=\overline{\Sigma}(<R)-\Sigma(R)=\gamma_t(R)\Sigma_c,
\end{equation}
where the critical surface density $\Sigma_c$ is 
\begin{equation}
\Sigma_c=\frac{c^2}{4\pi G} \, \frac{D_S}{D_L D_{LS} (1+z_L)^2}
\end{equation}
in comoving coordinates.  Here, $D_L$ and $D_S$ are the angular
diameter distances to the lens and source galaxies, and $D_{LS}$ is
the angular diameter distance between the lens and source galaxies.
The right side of Eq.~\ref{eqn:ds} shows how 
$\ds(R)$ is related to the observable $\gamma_t$, the mean tangential shear of source
galaxies behind the lens galaxies.  This equation is true in the weak
lensing limit ($\gamma_t \ll 1$) in the presence of an axisymmetric
mass distribution, which is achieved by our measurement procedure.

The measurement is performed by stacking lens galaxies based on their
luminosities.  The stacking procedure allows us to achieve sufficient
source number density to measure the shear signal, and therefore
$\ds$, of the composite object.  To do so, we measure the tangential
shear of all source galaxies in annular bins around the lens positions.
The details of the signal computation are described in appendix
\ref{app:signal}.  We are implicitly assuming
that on the scales used for our measurement, the shearing of
background galaxies is caused only by the lens halos, and not by other
nearby halos or infalling dark matter (in the language of the halo
model, we neglect the two-halo term).  For the few lenses that may be
satellites despite our selection criteria designed to avoid them, we
also neglect the contribution from the host halo.  Furthermore, we 
neglect the radial window that enters the line-of-sight
projection in Eq.~\ref{eqn:proj}, given that the window is far broader
than the characteristic scale of our lenses.

To generate model lensing signals that can be compared against the
measurements, we fit the two-parameter NFW profile to the lensing
data.  The spherical NFW density profile is defined as
\begin{align}\label{eqn:nfw}
\rho_\mathrm{NFW}(r)=\frac{\rho_s}{(r/r_s)(1+r/r_s)^2}.
\end{align}
We can express it in terms of two parameters, the mass $M$ within which
the average density is equal to some overdensity (here, 200 times overdense
compared to the mean background density $\bar{\rho}$), and the
concentration, which is the ratio of the virial radius to the scale
radius $c= r_{\rm vir}/r_s$.  $M$ and $c$ can be
related to $\rho_s$ and $r_s$ via
\begin{align}\label{eqn:200rho}
M=\frac{4 \pi}{3} r_{\rm vir}^3 (200\bar{\rho}) =4\pi \int_0^{r_{\rm vir}} \rho(r)\,  r^2 \rmd r \\
\bar{\rho}=\Omega_m \rho_c=\Omega_m\frac{3H^2}{8\pi G}.
\end{align}
Here $\Omega_m$ is the matter energy density today relative to the critical 
density, and $H$ is the Hubble parameter. 
Our model signal can then be generated for a given value of $M$ and
$c$ by projecting the profile along the line of sight
(Eq.~\ref{eqn:proj}) and integrating to obtain $\ds$
(Eq.~\ref{eqn:ds}).  

Our use of a spherical NFW profile is justified because of the large
number of lenses stacked, as verified by \cite{2005MNRAS.362.1451M}.
We use the NFW density profile rather than the Einasto profile,
\begin{equation}
\rho_\mathrm{Einasto}(r) =\rho_s e^{(-2/\alpha)[(r/r_s)^{\alpha}-1]}, 
\end{equation}
but note that
the two profiles are quite similar on weak lensing scales, though they
are substantially different at smaller scales.  
One consequence of this choice is
that the estimate of the concentration may potentially be
systematically biased in a way that depends on the innermost radius
used in the fit.  The impact of a systematic offset in concentration
is discussed in section \ref{sec:se}.  However, as shown in
\cite{2008JCAP...08..006M}, fitting the galaxy cluster weak lensing
profile on these scales for a concentration gave the same result to
within several per cent when using an NFW profile and an Einasto
profile.

The fitting procedure involves two steps.  In the first step, for each
luminosity bin we fit the profile to the observed $\ds$ using the
non-linear least-squares algorithm of Levenberg-Marquardt.  Here, we
use all data from $70<R<2000$ $h_{70}^{-1}$kpc even
though this may include scales on which lensing from large-scale
structure is important.  Our choice of minimum radius was motivated by
the fact that ellipticity measurements of source galaxies closer than
$R<70 \, h_{70}^{-1}\rm{kpc}$ to the lens center may be unreliable due to
confusion of light from the lens, and smaller source counts.  The
resulting profile with $(M,c)$ for each luminosity bin is used to
estimate a virial radius for each bin.  We then fit again, truncating
the data at that virial radius to avoid significant contributions to $\ds$ from
material outside the lens halo (two-halo term).  This
procedure results
in fitting fewer radial bins for the faintest sample.  The brightest and 
intermediate bins are fit in the interval $70<R<850$ and the faintest 
bin in the interval $70<R<570$ $h_{70}^{-1}$kpc. 

\begin{figure}
\begin{center}
\resizebox{3.45 in}{!}{\includegraphics{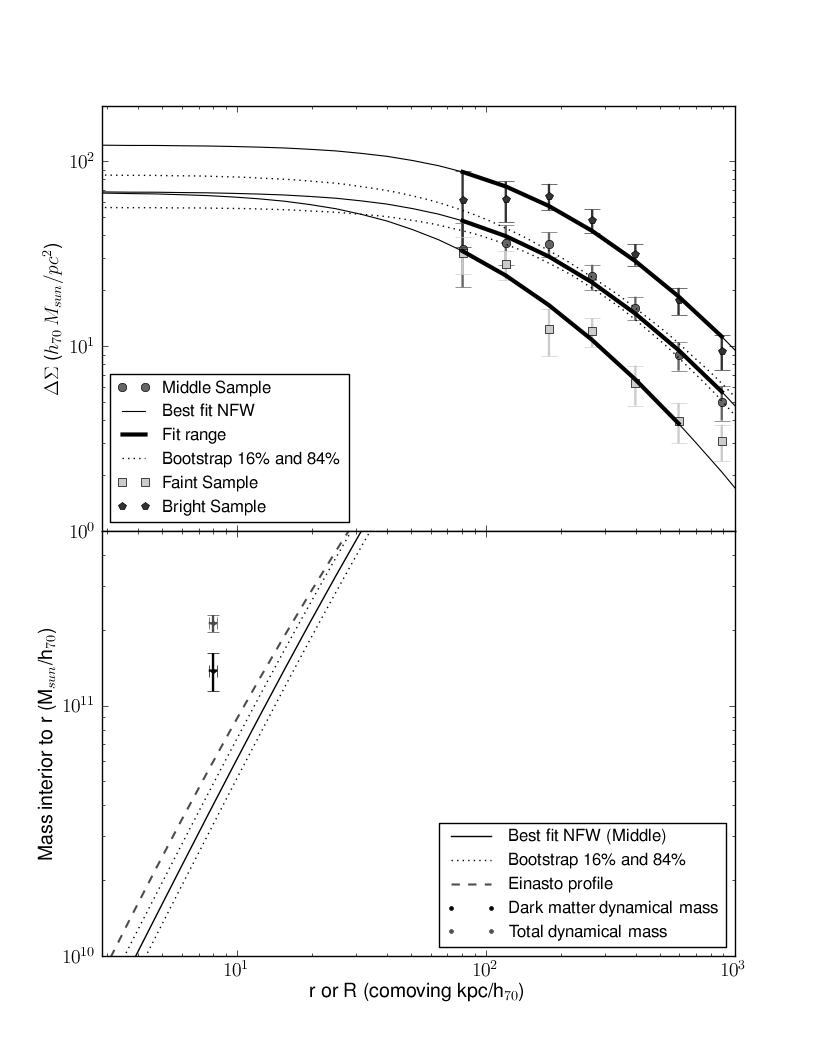}}
\end{center}
\caption{{\bf Top:} Points are the weak lensing measurement.  
Solid lines show the differential surface density for the best fit NFW model.  Dotted lines show 
16 and 84 per cent contours from fits to bootstrap resampled datasets. {\bf Bottom:}
 Solid lines show the integrated mass interior to $r$, and dotted
lines show a bootstrap estimate for error in NFW.  The dashed curve shows the result for 
an Einasto profile for comparison.
The upper point shows the total dynamical mass, while the lower point shows the dark matter contribution 
to the dynamical mass after subtracting the estimated contribution
from stellar mass.  
}
\label{fig:lnfw}
\end{figure}

The lensing result and the NFW fit are shown in the top panel of
Fig.~\ref{fig:lnfw}.  The heavy solid line shows $\ds$ for the best fit NFW model,
while the thin solid line shows the extrapolation of the NFW result to
small radii.  The points show the lensing measurements.  The dashed
lines shows a 68 per cent confidence level estimate for the error on the NFW 
result for the middle sample.  The errors are derived by bootstrap resampling 200
regions on the sky 1000 times, and 
recomputing the fit for each resampled dataset. The curves show the 16th and 84th 
percentile values of $\ds(R)$ determined by rank-ordering the fits to the bootstrapped datasets. 

The bottom panel of Fig.~\ref{fig:lnfw} shows the (3D) integrated mass interior to the comoving 
radius $r$ the NFW model of the middle sample shown in the top panel.  For comparison, we also plot the enclosed
mass for the Einasto profile, using the same mass and concentration parameters, and fixing the shape 
parameter following \cite{2008MNRAS.387..536G}. 
The horizontal scales in the two panels are the same.  The total dynamical 
mass (upper point) and the dark matter contribution to the dynamical
mass (lower point) are plotted
at the position of the 2D de Vaucouleurs half light radius  of the lens galaxies.  The half light radius 
and dynamical masses shown are the lensing weighted mean value for the lenses in this luminosity bin.  
The lensing weights, which can be found in appendix \ref{app:signal},
are assigned to each lens-source pair, and include redshift information and error on the shape
measurement of the source.  
To deduce the dark matter contribution to the 
dynamical mass, we subtract 42 per cent of the total stellar mass (the 
fraction of stars enclosed within a 3D radius corresponding to the 2D
half light radius) from the total dynamical mass. 
The error bar on $R_{\rm deV}$ comes from the measurement error.  The
error bar on the total dynamical mass is estimated to be $\sim 10$ per cent, from 
variations in the input modeling from Eqn. \ref{eqn:dym2} and from differences 
between $R_{\rm deV}$ and the portion of the galaxy covered by the SDSS fiber. 
The error bar on the dark matter dynamical mass comes from combining this systematic
error in the total dynamical mass with measurement errors in the stellar masses. 

Although the NFW curve has been extrapolated inward 
by more than an order of magnitude in transverse separation, it 
is noteworthy that the  total dynamical mass 
in the interior of the galaxy
exceeds the NFW prediction by eleven standard deviations.   
Even when the contribution of stars has been accounted 
for, there appears to be over a factor of two more dark matter 
mass in the central region of the galaxy 
(four standard deviations) than what is seen in halos from 
N-body simulations of large scale structure. 

This result is consistent with earlier attempts to observe the density profile of elliptical galaxies.
\cite{2007ApJ...671.1568J} used stellar dynamical and strong lensing
measurements of 22 early-type galaxies from the Sloan Lens ACS Survey
(SLACS) to constrain the total matter profile near the effective
radius.  These data alone are not enough to distinguish between a
model with and a model without adiabatic contraction; however, when
they include weak lensing constraints on $100$\hkpc\ scales from
\cite{2007ApJ...667..176G}, they find that the models with adiabatic
contraction are strongly favored.  For their analysis, they used the
\cite{1986ApJ...301...27B} model for adiabatic contraction, which
predicts a more significant effect than the \cite{2004ApJ...616...16G}
model used in our work.

\subsection{Comparison with adiabatic contraction model}\label{sec:ac}
\begin{figure}
\begin{center}
\resizebox{3.8 in}{!}{\includegraphics{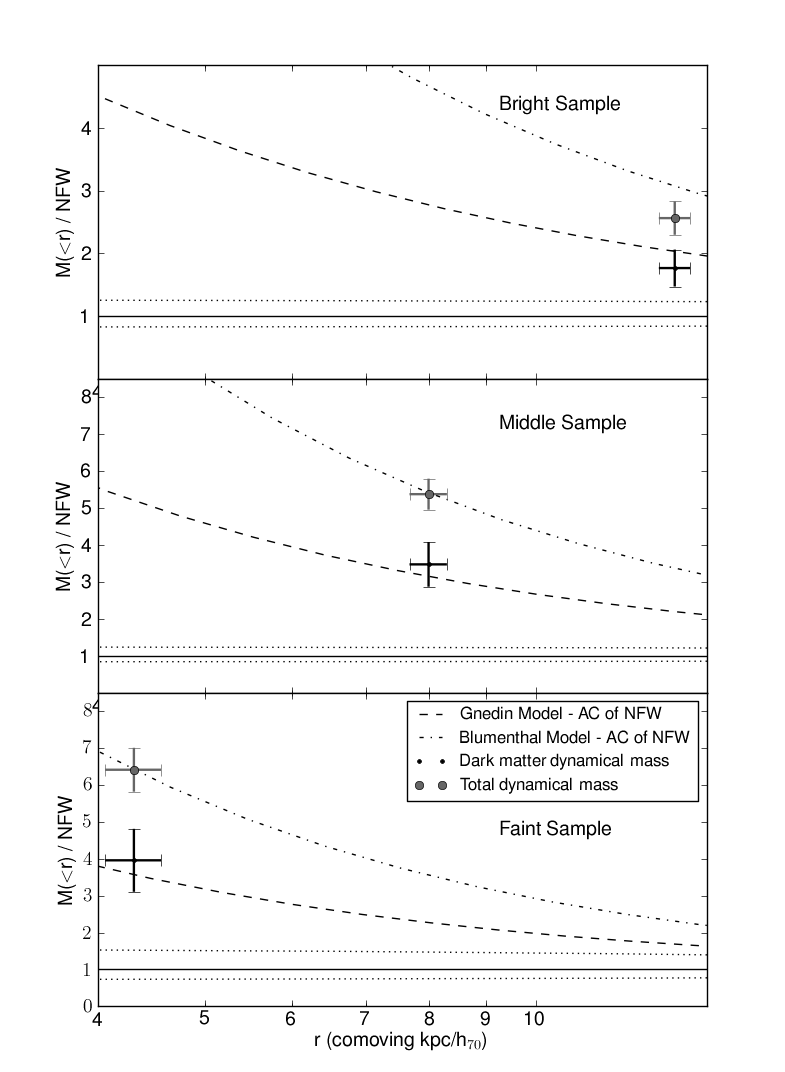}}
\end{center}
\caption{All curves and points are defined in the same convention as
  in the bottom panel of Fig.~\ref{fig:lnfw}, with the addition of a
  dashed line showing the predictions including AC.  The non-AC
NFW model has been divided out (solid line at 1)
so as to present the results on a linear scale.  The (top, middle,
bottom) panel shows the (brightest, intermediate, faintest) lens
sample.}
\label{fig:flat}
\end{figure}

It is well known that baryonic physics is expected to affect the form of the dark matter profile in 
galaxy halos, so it is not entirely surprising that the profile departs from N-body predictions 
in a region where the mass is largely made up of stars.   Exactly how the baryons affect 
the dark matter distribution, and whether their effect is transient or enduring 
is however a matter of some theoretical debate.  On the one hand, the
dark matter is expected to fall into the cuspy deep gravitational potentials that are created when baryons
cool and condense to form galaxies, a process known as adiabatic contraction (AC) that further 
deepens the central potential.  In the simulations of \cite{2008ApJ...672...19R}, these steeper potentials are seen to persist, and the impact
of the baryons is evident on the matter power spectrum, principally
through the change in the dark matter profiles inside the virial radii
of halos, particularly around cluster masses of $10^{14}h^{-1}M_{\odot}$.  On the other 
hand, there is some debate as to whether spiky central cusps will survive the onslaught of subsequent major and minor merger activity 
expected in hierarchical models of structure formation.   In the 
simulations of \cite{2009ApJ...697L..38J}, adiabatic contraction is seen initially, but is transformed to a cored central 
potential through gravitational heating from mergers.  To date, there have been very
few observational studies that can probe the dark matter halo
over the large range of scales required to test the AC hypothesis.   We make an attempt here 
by comparing the AC model of \cite{2004ApJ...616...16G} to the dynamical mass data, shown in Fig.~\ref{fig:flat}.

Each of the panels in Fig.~\ref{fig:flat} is analogous to the bottom panel of Fig. \ref{fig:lnfw}, but each quantity is now  divided by the best fit NFW relation (solid curve in Fig.~\ref{fig:lnfw}) so as to 
present the results on a 
linear scale.  Note that the best fit NFW is different for each
panel. 
The 3d stellar fractions of the total mass at the half light radius
are $f_*=31$, 35 and $38h_{70}^{-1}$ per cent (from bright to faint).
The curves in Fig.~\ref{fig:flat} show the AC models of
\cite{2004ApJ...616...16G} (dashed) and \cite{1986ApJ...301...27B} (dot-dashed).
These are computed using the publicly available {\sc contra} code (\citealt{2004ApJ...616...16G}).  
To model AC, we use the mass and concentration from the weak lensing measurements, and make a few reasonable assumptions.  We assume that for these elliptical lens galaxies all of the gas
has been converted into stars (\citealt{2005RSPTA.363.2693R},
\citealt{2006MNRAS.371..157M}, \citealt{1998ApJ...503..518F}), so that the baryon fraction is given by
the stellar mass divided by the best-fitting dark matter halo mass.  This assumption is robust for the 
fainter bins but may begin to break down for the brightest bin, for
which the mass approaches the group scale.
We also assume the baryons ultimately arrange 
themselves into a S\'ersic profile with $n=4$, i.e. a de Vaucouleurs
profile, with scale radius $R_{\rm deV}$ converted as in the appendix of \cite{2004NewA....9..329P}.

It is remarkable that in all three luminosity bins, the model NFW
profile with adiabatic contraction agrees with the
inferred contribution of dark matter to the total dynamical mass.  This AC model leads to enhancements of
the dark matter contribution to the dynamical mass that are
approximately factors of $4$, $3.5$, and $2$ for the faint, middle, and
bright samples respectively.  The disagreement of the data with the
extrapolated pure NFW profile (i.e., without AC) is significant at the $3.5\sigma$,
$4\sigma$, and $2.5\sigma$ level, respectively.  The apparent agreement between the 
total dynamical mass and the \cite{1986ApJ...301...27B} model of AC is a 
coincidence; for this to be meaningful, none of the mass could be in stars.  

\subsection{Potential systematic errors}\label{sec:se}
In this section, we discuss a number of systematic uncertainties present in this measurement that make the very close agreement
between the AC model and the measurement mildly surprising.  

\subsubsection{Initial mass function}
The most significant concern for potential
systematic error lies in the determination of the stellar mass.  The
stellar masses are computed assuming a Kroupa Initial Mass Function
\citep[IMF, ][]{2001MNRAS.322..231K}, which is calibrated by measuring the mass function of stars in the solar 
neighborhood.  While it is expected that some level of churning will bring stars from orbits in all parts of the galaxy close to the solar neighborhood, the degree of churning is very uncertain.  More importantly, galaxies of different ages and types have different metallicities, and the shape of the IMF may well 
depend on metallicity.  In \cite{2003MNRAS.341...33K}, 
the sensitivity of the stellar masses to the choice of IMF was 
investigated, and it was shown that assuming a Salpeter \citep{1955ApJ...121..161S} IMF with a lower mass cutoff at $0.1 M_{\odot} $ systematically increased the stellar masses by approximately a factor of 2.  However, the amount of increase in the stellar masses will quantitatively depend on how the low mass end of the (divergent) Salpeter IMF is regularized.  It is interesting to contemplate that aside from the question of AC, this measurement
of the dynamical mass can be thought of as a constraint on the IMF, in the sense that the stellar mass
cannot exceed the total dynamical mass.  

To fully explain the observed dynamical masses with no adiabatic contraction of the dark matter, the
stellar masses would have to increase by a factor of 1.9, 2.3 and 2.2 for the bright, middle and 
faint samples.  However,
\cite{2006MNRAS.366.1126C} present a challenge to this alternative
  explanation.  They use detailed integral field spectroscopy
  observations to infer dynamical masses within the effective radius
  for 25 nearby E/S0 galaxies, and compare them with the predicted stellar
  masses from stellar synthesis models.  They find
  that for some of the galaxies, the predicted stellar masses with a
  Salpeter IMF exceed the dynamical
  masses, which suggests that if we require the same IMF for each
  galaxy, then the Salpeter IMF is too bottom-heavy.  In contrast, 
  the Kroupa IMF gives stellar
  masses that are 30 per cent lower
  than with the Salpeter IMF (for their choice of cutoff mass), causing all the 
  stellar masses to be less than or equal to the inferred
  dynamical masses.  Thus, the conclusion that the IMF is very
  bottom-heavy (instead of accepting
  the AC hypothesis) may be difficult to reconcile with these IFU observations
  of elliptical and S0 galaxies. 

\subsubsection{Other stellar mass uncertainties}

There are other systematic effects in the stellar mass estimates that
are expected to be subdominant to the IMF uncertainties.  For example,
the effects of aperture bias (the fact that the $M_*/L$ are derived from the region
of the galaxy covered by the fiber, not the whole galaxy) are estimated
to be smaller than that of IMF uncertainties.  Also, aperture bias is more problematic for galaxies
with both differently-coloured bulge and 
disk contributions than for the pure ellipticals of which our sample
is composed.  See
\cite{2003MNRAS.341...33K} for a more full discussion of all stellar mass-related systematics.

\subsubsection{Initial conditions}

\begin{figure}
\begin{center}
\resizebox{3.6 in}{!}{\includegraphics{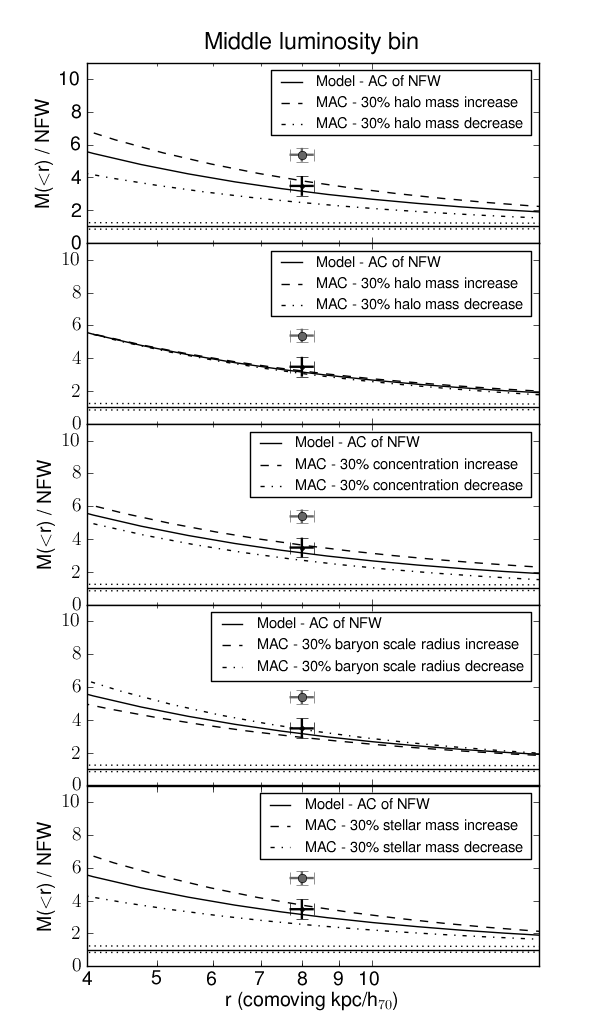}}
\end{center}
\caption{Each panel shows AC of the best fit NFW model (Solid curve), the total 
dynamical mass (upper grey point) and the dark matter contribution to the dynamical mass (lower
black point).  The dashed [dot-dashed] curve shows the impact on AC of increasing [decreasing] 
the halo mass (upper), concentration (upper middle), scale radius of the baryons (lower middle), and stellar mass (lower).  All results have been 
divided by the non-AC best fit NFW model.}
\label{fig:var}
\end{figure}

Another systematic uncertainty relates to the conditions under which
adiabatic contraction initially occurred.  The process of AC happens while the baryonic component is still in gaseous form, whereas the measurement is being 
made on galaxies that have converted nearly all of the gas into stars.  
Presumably several (major or minor) mergers 
have occurred in the interim, whose qualitative effect will be to raise the mass and lower the concentration of the halo.   
If the baryon fraction of the merging objects differs from that of the parent
halo, the baryon fraction will also have evolved since the time AC was occurring.   To quantify 
how sensitive the AC result is to changes in the input parameters, we show  
in Fig.~\ref{fig:var} the effects of a 30 per cent variation in halo 
mass, concentration, scale radius of the baryon component,  
and baryon fraction.  

The {\sc contra} code accepts the baryon mass fraction as an input, and 
returns the pre-contracted and post-contracted 
curves in units where the mass has been scaled to the NFW virial mass, which we measure in the
lensing analysis.  We define the baryon fraction as $f_b=M_*/M$
(assuming that all the gas has been
converted to stars)\footnote{Please note this is distinct from $f_*$ in Table \ref{tab:results}, which 
indicates the fraction of the total dynamical mass that is in stars
within a sphere of radius equal to the half light radius $R_{\rm deV}$.}.  The top two panels of Fig. 8 show the result of varying the halo mass $M$, in
one case fixing the baryon fraction (top) and in the other case allowing the variation of $M$ 
to affect the baryon fraction (second panel).  In the top panel, changing the mass simply scales the 
curves up and down.  The second panel shows that the change from scaling the 
mass in the top panel is nearly exactly offset at these radii 
by the effect of decreasing the baryon fraction.  This result can be understood by comparing the top 
panel with the bottom panel (where only the stellar mass is scaled).  A decrease in stellar mass 
implies a decrease in the baryon fraction, and since the deviation from the fiducial model is 
roughly the same size in the top and bottom panels, the effects cancel one another out in the 
second panel.  One other noteworthy feature in Fig. 8 is that the effect of increasing the baryon 
scale radius changes the level of AC in the opposite sense than the other 4 curves.  
We conclude that there is a modest sensitivity in the model to these
quantities that could change the prediction at a level comparable to the measurement error of the 
dynamical mass (indeed this is the reason we selected a 30 per cent variation).   This sensitivity is 
not so large as to invalidate the result for reasonable (tens of per cent) shifts in these parameters.  

\subsubsection{Dynamical mass assumptions}

It is quite probable that the lens sample violates the assumption that  the radial 
anisotropy parameter $\beta$ is zero.  It has already been shown in 
\cite{2004NewA....9..329P} that the measured
velocity dispersion is insensitive to the value of 
the radial anisotropy parameter.  However it is also worth noting that for this type of galaxy, the 
anisotropy is much more likely to be radial rather than tangential \citep{2007MNRAS.379..418C}.  
The impact of assuming no radial anisotropy for such objects will be to systematically underestimate
the dynamical mass within the half light radius.  A similar systematic effect that causes the dynamical
mass to be underestimated is if the galaxies have some moderate rotational support.  Both of these
effects strengthen the conclusion that there appears to
be more mass than predicted by simple extrapolation of the NFW profile; our dynamical mass measurements can be considered a lower bound.

\subsubsection{Spread in the properties of the lens subsamples}

Finally, our analysis uses relatively broad luminosity bins (because
of the need for large lens samples to measure the lensing signal at
reasonable $S/N$).  Unfortunately, this means that a large amount of
averaging takes place, which could muddy the interpretation of the
results.  We have attempted to ameliorate this effect by computing
lensing-weighted average quantities (stellar masses, dynamical masses,
half-light radii, and so on) to compare against the best-fitting dark
matter halo mass.  This procedure should help make the different kinds
of measurements more comparable.  However, \cite{2005MNRAS.362.1451M}
show that for lens samples with halo mass distributions that are more
than a factor of five wide, the best-fitting halo masses lie
somewhere above the median and below the mean, an effect that can be
as large as several tens of per cent in extreme cases.  Since Fig.~\ref{fig:var} shows that
changing the halo mass and other properties by 30 per cent does not
change the AC predictions by more than about $1\sigma$, we anticipate
that our neglect of the bin width (which is unknown, in the case of
the halo mass) will not substantially alter our
conclusion that the AC model is consistent with our findings.

\subsection{Splitting along the size-luminosity relation}
\begin{figure}
\begin{center}
\resizebox{3.5 in}{!}{\includegraphics{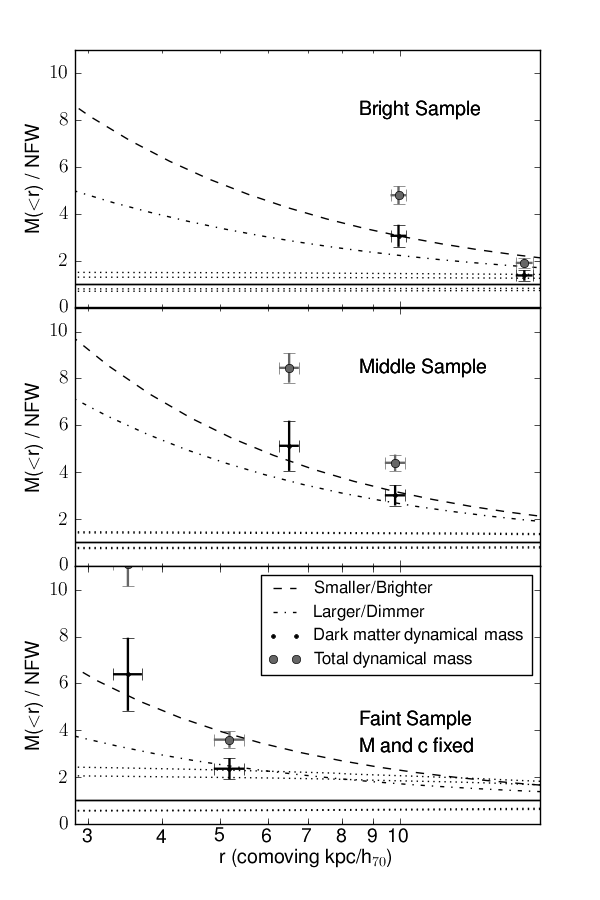}}
\end{center}
\caption{Three luminosity bins are shown from brightest 
(top) to dimmest (bottom panel).   In each panel, curves on the top (bottom) show the AC 
results for lens subsamples that are smaller and brighter
(larger and dimmer) 
the mean size luminosity relation.  The solid lines with dotted
error estimates show the enclosed mass for the NFW 
model that best fits the weak lensing data.   The total observed dynamical mass (upper grey 
point) and the dark matter contribution to the dynamical mass (lower black point) plotted at 
the position of the half light radius.  Points on the left (right) are the smaller/brighter 
(larger/dimmer) sample.   Because the mass and concentration cannot be independently 
constrained for the two subsets of the faintest sample, $M$ and $c$ are fixed to the values of the composite 
faint sample (top level of Table \ref{tab:results}).
}
\label{fig:sizelum}
\end{figure}

It is interesting to ask if the level of excess mass in the galaxy interior correlates 
with other galaxy properties.  There is evidence that a galaxy's environment and
merger history directly impact the concentration and relative distributions of baryons 
and dark matter (\citealt{2009arXiv0902.2477A} and references therein).  
It has also been suggested that multiple major or minor 
mergers may erase some of the effects of adiabatic contraction (\citealt{2009ApJ...697L..38J}).  
Since elliptical galaxies lie tightly on a fundamental plane, it 
would also be interesting if small departures correlate strongly with the form of the galaxy's 
dark matter profile.
Fortunately, there is enough signal in the data to support one further level of 
subdivision in the analysis.  We have opted to examine these questions by dividing 
each luminosity bin along the mean size-luminosity relation, which is a projection of the
fundamental plane.  The mean relation in our sample is
\begin{align}
\log(R_{\rm deV})=0.722+0.598*((-20.44-M_r)/2.5)
\end{align}
where $R_{\rm deV}$ is in units of $h_{70}^{-1}$ kpc.
The results are presented in Table \ref{tab:results} and Fig.~\ref{fig:sizelum}.  
The upper set of curves (dashed lines) show the departure from NFW (solid line at 1) due to AC for the
smaller/brighter sample.  The lower set of curves (dot-dashed lines) show AC for the larger/dimmer sample.  The points 
on the left (right) correspond to the smaller/brighter (larger/dimmer) sample.   Table \ref{tab:results} 
indicates that for the faint sample, we are unable to simultaneously constrain $M$ and $c$ for these 
subsamples divided along the mean size-luminosity relation.  Therefore, in the bottom panel of Fig. 
\ref{fig:sizelum}, we adopt the $M$ and $c$ from the composite sample
(row 3 of Table \ref{tab:results}). As a result, for the top two
panels, the AC curves and data points for the two subsamples have been divided out by a
different non-contracted NFW model; on the bottom panel, by the same model.

Table \ref{tab:results} shows that the halo masses and half-light radii of the smaller/brighter (S/B) sample
are smaller than those of the larger/dimmer (L/D) sample, but the stellar masses and concentrations 
of the two populations are nearly the same.    
Fig. \ref{fig:sizelum} indicates that the total dynamical mass at the half light radius is 
much more enhanced above the NFW prediction for the S/B sample than for the 
L/D sample.  There are competing effects driving this trend; it is helpful to refer to 
Fig. \ref{fig:var} in deciphering the result.  Since the stellar masses are equivalent, the smaller 
halo masses for the S/B sample imply a larger baryon fraction $f_b$ than in the L/D sample.  Just
as in panel two of Fig. \ref{fig:var}, the increase in mass is offset by the decrease in the baryon 
fraction in comparing the S/B to the L/D sample.  The concentrations of the two samples do not 
change very much, but the significant change in the baryon scale radius $R_{\rm deV}$ drives
the overall trend.  Since the S/B sample has a larger baryon fraction, and more of the baryons are 
concentrated toward the center of the halo, the effects of AC are more
pronounced than in the
L/D sample, even though the L/D sample has higher overall halo mass.  
There is also a higher stellar mass fraction $f_*=(0.42 M_*)/M_{\rm dy}^{Tot}$
at the half light radius for the 
S/B sample than for the L/D, although part of this effect
could be accounted for by non-homologous profiles (P. v. Dokkum, priv comm).  

An alternate interpretation of this result might be that the stellar IMF for
the S/B sample has more mass in small faint stars than the IMF for the L/D sample 
(see the discussion of systematic error in Section~\ref{sec:lensing} above), or potentially some 
combination of these two effects.  The IMF interpretation requires us
to discard the idea of a universal IMF for all elliptical galaxies.

Since it is likely that the S/B sample had its stellar population in place 
at earlier times, and that the L/D has potentially experienced merger 
activity more often or more recently, it would be very interesting in the future to correlate 
the level of AC with other indicators of age or merger activity.  
It is also interesting to note that the L/D subsamples within each
luminosity bin have consistently larger halo masses than the S/B
subsamples, despite the fact that their stellar masses are very
similar.  This result suggests that different star formation histories
(that led to the different appearance of the stellar components at
fixed stellar mass) also correlate with different accretion histories for
the dark matter halos.

\subsection{Comparison with previous work}

Here we discuss how our results compare with previous, related
observations.  We first compare with \cite{2004NewA....9..329P}, which
is relevant due to the similar sample definition and methodology.
Since that work did not have any observational method of constraining
the profile on large scales, we simply compare against their
determination of the 3d stellar mass fraction $f_*$ within the
half-light radius.  The relevant comparison is against figure 8 of
that paper, which shows $M_\mathrm{dy}/M_*$ (i.e., our 
$f_*^{-1}$) as a function of the galaxy apparent size, for galaxy
subsamples split by stellar mass.  For our three relatively broad
luminosity subsamples, we have typical values of half-light radius
range from 4 to 13$h_{70}^{-1}$kpc and typical $1/f_*$ ranging from 2.6 to 3.2$h_{70}$;
\cite{2004NewA....9..329P} find values of $1/f_*$ that range from 2--4
(depending on the stellar mass, in narrower bins than ours), which
is consistent with our findings. 

The comparison with \cite{2007ApJ...667..176G} is less trivial due to
the very different methodology.  In that paper, they model the strong
and stacked weak lensing signals for SLACS lenses jointly using the sum of an uncontracted NFW
and a stellar Hernquist profile with a free $M_*/L$ (without any
stellar mass estimates from stellar population synthesis methods).
With this procedure, 
they quote a 3d stellar mass fraction within the effective
radius of $f_*=0.73$ (independent of $h$ because of how the modeling
is done).  However, \cite{2007ApJ...671.1568J} model the
same SLACS sample, and find that if they allow for the possibility of
adiabatic contraction, the resulting $M_*/L$ is 70 per
cent of the value when adiabatic contraction is ignored (see their
figure 6).   This decrease
results from a degeneracy in the fits: an
uncontracted DM profile requires more mass to be attributed to stars
than a contracted DM profile, in order to account for the total mass.  Furthermore, they find that the fit
with AC (and lower value of $M_*/L$) is preferred once they include
the weak lensing in the modeling.  Consequently, the \cite{2007ApJ...671.1568J} results
suggest that this value of $f_*$
from \cite{2007ApJ...667..176G} should be reduced to $\sim 0.5$, which is
higher than our values of $0.31$--$0.38$, but not very significantly 
once the different sample selections (strong lensing versus optical)
and redshift ranges are taken into account. 

Finally, we compare against the results from \cite{2006MNRAS.366.1126C}, who find that for 25 SAURON E/S0 
galaxies, the assumption of a Kroupa IMF yields $f_* \sim 0.7$ (median
value, independent of $h$).  While this number is significantly 
higher than ours, the selection criteria and redshift range of the galaxies are significantly different, so it is difficult to 
derive any conclusions from this comparison.

\section{Conclusions}\label{sec:conclusions}
We have combined weak lensing and velocity dispersion observations to study the dark matter 
profiles of elliptical galaxies from the SDSS.  The radial profile of the weak lensing 
shear is consistent with that expected from an NFW halo profile.  We have fit two 
NFW parameters to the weak lensing data (mass and concentration), and extrapolated the 
profile inward to smaller radii not accessible to weak gravitational lensing.  We have deduced the 
dynamical mass at these smaller radii by measuring the velocity dispersion of the stars within the 
half light radius of the galaxy.   We compare the measured dynamical mass to the extrapolation of the NFW profile, and find that there is a significant excess of mass in the interior.   Using estimates of the stellar mass
of the galaxy, and assuming that all non-stellar mass is dark matter, we find that the dark matter 
contribution to the dynamical mass is still far in excess of the NFW
prediction.  This result is in support of the model of adiabatic
contraction (AC), which predicts that the cooling and condensation 
of baryonic material will deepen the gravitational potential well of
the galaxy, and pull the dark matter towards the halo center.   These results suggest that the effects of AC are stable to the subsequent bombardment of 
major and minor mergers suffered by these objects since the time their gas cooled.  

We compare the observation to a theoretical model of adiabatic
contraction \citep{2004ApJ...616...16G} and find good 
agreement between the two.  However, we acknowledge that systematic uncertainties in the 
determination of the stellar mass make it difficult to prove that the excess mass at the half light radius
is indeed dark matter.  An alternative interpretation may be that the excess mass is comprised of some other dark baryonic form, for example it may indicate an IMF with
more of the mass concentrated in dim, low mass stars relative to the Kroupa IMF used in our stellar mass estimates.  In order to
fully explain the observation without adiabatic contraction, the stellar masses would need to
increase by a factor of two.  A change of this magnitude in the stellar masses would be difficult
to reconcile with recent results from the SAURON collaboration.  
They conclude that a 30 per cent 
increase in stellar mass (from masses obtained with the Kroupa IMF) is 
inconsistent with their dynamical data derived from integral-field spectroscopy. 

Finally, we divide the sample of galaxies along the mean size-luminosity relation and show that the 
scatter from the mean relation is related to the underlying properties of the dark matter profile.  We
find that due to a higher baryon fraction with more of the baryons concentrated toward the center, the 
smaller/brighter subsample experiences more lasting effects of adiabatic contraction than the larger/dimmer subsample.  If adiabatic contraction is not the explanation, the most likely alternative is 
different IMFs for the two subsamples.   Also, we find that the two subsamples have comparable stellar masses but
different halo masses, suggesting that stellar mass does not trace
halo mass, possibly due to different formation histories that result in
differing dark and stellar components (with larger halo mass being
associated with a dimmer, less concentrated stellar component).

Ultimately, regardless of whether the excess mass is dark matter or baryonic, this set of observations
indicates that the total mass at the half light radius far exceeds what would naively be expected from an
NFW or Einasto profile.
Contemporary models of galaxy formation and 
evolution must accommodate this relatively high ratio of dynamical mass in the interior to total halo 
mass in early type elliptical galaxies. 

\section*{Acknowledgments}

The authors thank Glenn van de Ven, Scott Tremaine, and Charlie Conroy
for inspiring  and helpful 
conversations.  NP thanks Pieter
van Dokkum for useful discussions.  A.E.S. is supported by the Corning Glassworks fellowship.
R.M.'s work on this project was supported by NASA
through Hubble Fellowship grant \#HST-HF-01199.02-A, 
awarded by the
Space Telescope Science Institute, which is operated by the
Association of Universities for Research in Astronomy, Inc., for NASA, 
under contract NAS 5-26555. 

Funding for the SDSS and SDSS-II has been provided by the Alfred
P. Sloan Foundation, the Participating Institutions, the National
Science Foundation, the U.S. Department of Energy, the National
Aeronautics and Space Administration, the Japanese Monbukagakusho, the
Max Planck Society, and the Higher Education Funding Council for
England. The SDSS Web Site is {\em http://www.sdss.org/}. 

The SDSS is managed by the Astrophysical Research Consortium for the
Participating Institutions. The Participating Institutions are the
American Museum of Natural History, Astrophysical Institute Potsdam,
University of Basel, University of Cambridge, Case Western Reserve
University, University of Chicago, Drexel University, Fermilab, the
Institute for Advanced Study, the Japan Participation Group, Johns
Hopkins University, the Joint Institute for Nuclear Astrophysics, the
Kavli Institute for Particle Astrophysics and Cosmology, the Korean
Scientist Group, the Chinese Academy of Sciences (LAMOST), Los Alamos
National Laboratory, the Max-Planck-Institute for Astronomy (MPIA),
the Max-Planck-Institute for Astrophysics (MPA), New Mexico State
University, Ohio State University, University of Pittsburgh,
University of Portsmouth, Princeton University, the United States
Naval Observatory, and the University of Washington. 


\bibliographystyle{mn2e}
\bibliography{ellipticals,wlref}

\appendix

\section{Computation of the weak lensing signal}\label{app:signal}

We compute the lensing signal in logarithmic radial bins from 30
$h_{70}^{-1}$kpc to 2.9 $h_{70}^{-1}$Mpc
as a weighted summation over lens-source pairs in the bin, using the
following estimator:
\begin{equation}
\ds(R) = \frac{\sum_{ls} w_{ls} \gamma_t^{(ls)} \Sigma_c^{(ls)}}{2 \mathcal{R}\sum_{ls} w_{ls}}. 
\end{equation}
where the factor of $2\mathcal{R}$ arises due to the response of our
ellipticity definition to a shear.  $\mathcal{R}$, known as the ``shear
responsivity,'' is approximately $1-e_\mathrm{rms}^2\approx 0.87$
\citep{2002AJ....123..583B}. 

The weight factors $w_{ls}$ assigned to each pair include redshift
information and the error on the source shape measurement via
\begin{equation}
w_{ls} = \frac{\Sigma_c^{-2}}{\sigma_e^2 + \sigma_\mathrm{SN}^2}
\end{equation}
where $\sigma_\mathrm{SN}^2$, the intrinsic shape noise, was determined as a
function of magnitude in \cite{2005MNRAS.361.1287M}, fig. 3.  The factor of
$\Sigma_c^{-2}$ downweights pairs that are close in redshift, so that
we are weighting by the inverse variance of \ds.

The critical surface density, $\Sigma_c$, is computed in two different
ways, depending on the source sample.  For sources with $r<21$, we use
photometric redshifts, and require that the source photometric
redshift be greater than the lens redshift plus $0.1$.  The procedure
for correcting these $\Sigma_c$ estimates to account for photometric
redshift error is described in
\cite{2005MNRAS.361.1287M,2008MNRAS.386..781M}.   For sources with
$r>21$, we use a redshift distribution for the sources, so that for a
given lens redshift a single $\Sigma_c$ value is used for all sources
at $r>21$.  Finally, for the high-redshift LRG source sample used only
at transverse separations $<140h_{70}^{-1}$kpc, we use photometric
redshifts to assign $\Sigma_c$ for each lens-source pair.

There are several additional procedures that must be done when
computing the signal (for more detail, see
\citealt{2005MNRAS.361.1287M}).  First, the signal computed around
random points must be subtracted from the signal around real lenses to
eliminate contributions from systematic shear.  In practice, this
correction is negligible on the scales used for this measurement.
Second, the signal must be boosted, i.e. multiplied by $B(R) =
n(R)/n_{rand}(R)$, the ratio of the number density of sources around
real lenses relative to their number density around random points, in
order to account for dilution by sources that are physically
associated with lenses, and therefore not lensed.

In order to determine errors on the lensing signal, we divide the
survey area into 200 bootstrap subregions, and generate 1000
bootstrap-resampled datasets. 
\end{document}